\documentclass[conference]{IEEEtran}
\IEEEoverridecommandlockouts
\usepackage[T1]{fontenc}
\usepackage{cite}
\usepackage[cmex10]{amsmath}
\usepackage{amsthm}
\usepackage{amssymb}
\usepackage{bbm}
\usepackage{mathtools}
\usepackage[hidelinks]{hyperref}
\usepackage{algorithmic}
\usepackage{textcomp}
\usepackage[usenames, dvipsnames]{color}
\usepackage[linesnumbered,ruled]{algorithm2e}
\usepackage{subcaption}
\usepackage{graphicx}  
\usepackage{url}
\usepackage[usestackEOL]{stackengine}
\usepackage{verbatim}
\usepackage{balance}
\usepackage {tikz}
\tikzset{>=stealth}
\usepackage{xcolor,colortbl}
\definecolor{lightred}{RGB}{255, 240, 240}
\definecolor{lightblue}{RGB}{230, 250, 255}
\definecolor{lightgreen}{RGB}{240, 255, 242}
\definecolor{myred}{RGB}{220, 0, 0}
\definecolor{myblue}{RGB}{0, 17, 173}
\definecolor{mygreen}{RGB}{2, 117, 0}

\newcommand{\bs}[1]{\boldsymbol{#1}}
\newcommand{\mc}[1]{\mathcal{#1}}

\newcommand{\mb}[1]{\mathbf{#1}}
\newcommand{\mbb}[1]{\mathbb{#1}}
\newcommand{\mr}[1]{\mathrm{#1}}

\newcommand{\tr}{\operatorname{tr}}

\newcommand{\diag}[1]{\operatorname{diag}{#1}}

\newcommand{\lr}[1]{\langle #1 \rangle}

\begin{document}
	%
	\title{Variational Bayes Inference for Data Detection in Cell-Free Massive MIMO}
	
	\author{\IEEEauthorblockN{Ly V. Nguyen\IEEEauthorrefmark{1}, Hien~Quoc~Ngo\IEEEauthorrefmark{2}, Le-Nam~Tran\IEEEauthorrefmark{3}, A. Lee Swindlehurst\IEEEauthorrefmark{4}, and Duy H. N. Nguyen\IEEEauthorrefmark{5}}
		\IEEEauthorblockA{\IEEEauthorrefmark{1}Computational Science Research Center, San Diego State University, CA, USA}
		\IEEEauthorblockA{\IEEEauthorrefmark{2}Institute of Electronics, Communications, and Information Technology (ECIT), Queen’s University Belfast, UK}
		\IEEEauthorblockA{\IEEEauthorrefmark{3}School of Electrical and Electronic Engineering, University College Dublin, Ireland}
		\IEEEauthorblockA{\IEEEauthorrefmark{4}Department of Electrical Engineering and Computer Science, University of California, Irvine, CA, USA}
		\IEEEauthorblockA{\IEEEauthorrefmark{5}Department of Electrical and Computer Engineering,  San Diego State University, CA, USA}
		Email: vnguyen6@sdsu.edu, hien.ngo@qub.ac.uk, nam.tran@ucd.ie, swindle@uci.edu, duy.nguyen@sdsu.edu
	}
	
	\maketitle
	
	\begin{abstract}
	Cell-free massive MIMO is a promising technology for beyond-5G networks. Through the deployment of many cooperating access points (AP), the technology can significantly enhance user coverage and spectral efficiency compared to traditional cellular systems. Since the APs are distributed over a large area, the level of favorable propagation in cell-free massive MIMO is less than the one in colocated massive MIMO. As a result, the current linear processing schemes are not close to the optimal ones when the number of AP antennas is not very large. The aim of this paper is to develop nonlinear variational Bayes (VB) methods for data detection in cell-free massive MIMO systems. Contrary to existing work in the literature, which only attained point estimates of the transmit data symbols, the proposed methods aim to obtain the posterior distribution and the Bayes estimate of the data symbols. We develop the VB methods accordingly to the levels of cooperation among the APs. Simulation results show significant performance advantages of the developed VB methods over the linear processing techniques. 
	\end{abstract}
	

	\IEEEpeerreviewmaketitle

	\section{Introduction}
	\label{sec_introduction}

\begin{figure}[t]
		\centering
		\includegraphics[width=.8\linewidth]{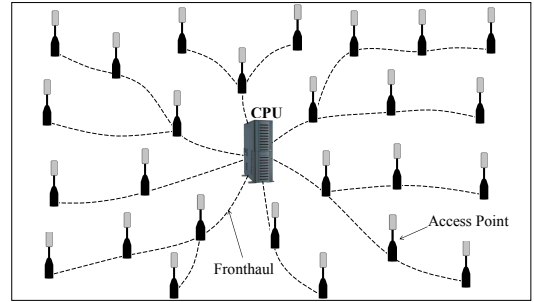}
		\caption{Diagram of a cell-free massive MIMO system with multiple distributed APs connected to a CPU.}
		\label{fig_cell_free_mMIMO_system}
	\end{figure}

Cell-free massive multiple-input multiple-output (MIMO) is considered as a promising technology for powering beyond-5G networks. The key idea of a cell-free massive MIMO system is to distributively deploy a large number of access points (APs) coherently serving all users in the system. As illustrated in Fig.~\ref{fig_cell_free_mMIMO_system}, the APs in a cell-free system can be randomly located all over the coverage area and are connected to one or several central processing units (CPUs). Due to this distributed deployment, any user is highly likely to be close to at least one AP. A cell-free system can effectively resolve the poor coverage issue in cell-edge areas of conventional cellular systems ~\cite{Ngo2017cell-free,Interdonato,Emil2020Scalable}. In addition, a cell-free system enables different levels of cooperation among the APs with certain levels of joint signal processing at the CPU, ranging from fully centralized processing (\emph{Level 4}), to partially distributed processing (\emph{Levels 3} and \emph{2}), and to a fully distributed processing (\emph{Level 1}) \cite{Emil2020Making}. Joint signal processing at the system's CPU allows a cell-free system to better address the inter-cell interference, which becomes more severe in cellular systems with small cell deployments. Therefore, cell-free massive MIMO systems can offer significant enhancements in user coverage and energy efficiency compared to traditional cellular systems~\cite{Ngo2017cell-free,Nayebi2017Precoding,Emil2020Making}. 
	
	The majority of existing research on uplink cell-free massive MIMO has focused on spectral and energy efficiency analysis with linear signal processing methods, such as maximum-ratio combining (MRC) \cite{Ngo2017cell-free}, zero-forcing (ZF) \cite{Ngo2017cell-free}, and linear minimum mean-squared error (LMMSE) \cite{Emil2020Making}. While such approaches have relatively low complexity, linear methods do not perform well in systems  with low level of favorable propagation (e.g. when the number of AP antennas is small or is not much larger than the number of UEs, or the channels are highly correlated). Nonlinear signal processing is thus a promising alternative approach that can offer higher spectral efficiency \cite{Emil2020Making} or lower bit error rate (BER) \cite{Song-TWC-2021}. The recent work in \cite{Song-TWC-2021} proposed a nonlinear optimization-based algorithm for joint channel estimation and data detection in cell-free massive MIMO. However, the approach in \cite{Song-TWC-2021} can only provide point estimates of the data symbols of interest. Different from these papers, the focus of this paper is on devising efficient algorithms to obtain Bayesian estimates of the data symbols. Unfortunately, realizing the exact posterior distributions of the data symbols is intractable, even in a conventional single cell MIMO system. We, therefore, develop variational Bayes (VB) inference methods for approximating intractable posterior distributions of data symbols, which are then used to detect the symbols. We investigate the VB methods for joint data detection with fully centralized processing at the CPU, as well as for distributed data detection at the APs. For fully centralized processing, we assume that full knowledge of the channel state information (CSI) is available at the CPU. Likewise, for distributed processing at each AP, we assume that CSI knowledge for the channel from the users to that AP is locally available. Simulation results show significant performance advantages of the developed VB methods over the LMMSE processing techniques in \cite{Emil2020Making}.

	
	
	\textit{Notation:} Upper-case and lower-case boldface letters denote matrices and column vectors, respectively. The transpose and conjugate transpose are denoted by $[\cdot]^T$ and $[\cdot]^H$, respectively. $\mc{CN}(\bs{\mu},\bs{\Sigma})$ represents a complex Gaussian random vector with mean $\bs{\mu}$ and covariance matrix $\mb{\Sigma}$; $\mc{CN}(\mb{x};\bs{\mu},\mb{\Sigma}) = \big(1/\big(\pi^K|\mb{\Sigma}|\big)\big)\mr{exp}\big(-(\mb{x}-\bs{\mu})^H\mb{\Sigma}^{-1}(\mb{x}-\bs{\mu})\big)$ denotes the probability distribution function (PDF) of a length-$K$ random vector $\mb{x}\sim \mc{CN}(\bs{\mu},\mb{\Sigma})$. $\mathbb{E}_{p(x)}[x]$ and $\mr{Var}_{p(x)}[x]$ are the mean and the variance of $x$ with respect to its distribution $p(x)$; $ \langle x\rangle$ and $\sigma_{x}^2$ denote the mean and variance of $x$ with respect to a variational distribution $q(x)$.
	
	\section{System Model}
	\label{sec_system_model_and_problem_formulation}

	We consider an uplink cell-free massive MIMO system with $L$ distributed APs, each equipped with $N$ antennas, serving $K$ randomly located single-antenna users. 
	It is assumed that $N\leq K\leq NL$. 
Denote $\mb{h}_{i\ell} \in \mbb{C}^{N}$ as the uplink channel from the $i$-th user and the $\ell$-th AP and $\mb{H}_{\ell} = [\mb{h}_{1\ell},\ldots,\mb{h}_{K\ell}]$. We assume a block Rayleigh fading scenario in which the channel $\mb{h}_{i\ell}$ remains constant for $T$ time slots and is normally distributed as $\mc{CN}(\mb{0},\beta_{i\ell} \mb{R}_{i\ell})$. Here,  $\beta_{i\ell}$ is the large-scale fading coefficient and $\mb{R}_{i\ell}$ is the normalized spatial correlation matrix whose diagonal elements equal to one. Due to the random user deployment, the large-scale fading coefficient $\beta_{i\ell}$ is different from one user to another user, resulting in a non-i.i.d. channel matrix $\mb{H}_{\ell}$. We assume that the channel vectors $\{\mb{h}_{i\ell}\}$ are independent of each other for each user-AP pair.
	
	Let $\mb{x}_t = [x_{1,t}, \ldots, x_{K,t}]^T$ be the transmitted symbol vector at time slot $t$, in which the transmitted symbol $x_{i,t}$ from the $i$-th user is drawn from a complex-valued discrete constellation $\mc{S}$ such that $\mathbb{E}[x_{i,t}] = 0$ and $\mathbb{E}[|x_{i,t}|^2] = \rho_i$. The prior distribution of $x_{i,t}$ is thus given by 
	\begin{eqnarray}\label{prior-x}
		p(x_{i,t}) = \sum_{a\in \mc{S}} p_a\delta(x_{i,t}-a),
	\end{eqnarray}
where $p_a$ corresponds to the known prior probability of the constellation point $a\in \mc{S}$.  The received signal vector $\mb{y}_{\ell,t} \in \mbb{C}^{N}$ at  the $\ell$-th AP can be modeled as
	\begin{equation}
	\mb{y}_{\ell,t} = \sum_{i=1}^{K} \mb{h}_{i\ell} x_{i,t} + \mb{n}_{\ell,t} = \mb{H}_{\ell}\mb{x}_t + \mb{n}_{\ell,t},
	\end{equation}
	where $\mb{n}_{\ell,t}$ is the noise vector whose elements are independent and identically distributed (i.i.d.) as $\mc{CN}(0,N_0)$. The interest of this paper is to obtain an estimated $\hat{\mb{x}}_t$ of $\mb{x}_t$ from multiple observed signal vectors $\mb{y}_{\ell,t}$'s across the $L$ distributed APs with minimum mean squared detection error $\mathbb{E}\big[\|\mb{x}_t-\hat{\mb{x}}_t\|^2\big]$. 
		
	\section{Four Levels of Cell-Free Massive MIMO Signal Processing Using LMMSE Filtering}
	To frame the discussion on the developed VB methods, we revisit the 4 levels of signal processing in cell-free systems using LMMSE filtering as studied in \cite{Emil2020Making}. Since the processing is based on a per time slot basis, without loss of generality, we drop the time index $t$.
	
	\subsection{Level 4: Fully Centralized Processing}
	At this level, the APs do not process their received signals. Instead, the received signals are forwarded to the CPU for fully centralized processing, including the data detection task. The signals forwarded from the $L$ APs can be stacked into
	\begin{equation}\label{eq_large_MIMO}
	\mb{y} = \mb{Hx} + \mb{n},
	\end{equation}
	where $\mb{y} = [\mb{y}_1^T, \ldots, \mb{y}_L^T]^T$, $\mb{H} = [\mb{H}_1^T, \ldots, \mb{H}_{L}^T]^T$, and $\mb{n} = [\mb{n}_1^T, \ldots, \mb{n}_L^T]^T$. The processing for cell-free massive MIMO in this level is similar to the processing at a conventional co-located MIMO receiver.  The CPU detects $\mb{x} = [x_1,\ldots,x_K]^T$ using the received signal vector $\mb{y}$ and the channel matrix $\mb{H}$. Among the linear detectors, the  LMMSE detector maximizes the signal-to-interference-and-noise ratio (SINR) and also achieves the best detection performance \cite{Emil2020Making}. With the full knowledge of $\mb{H}$, the LMMSE estimate $\hat{\mb{x}}$ is formed as
	\begin{eqnarray} 
		\hat{\mb{x}} = 
	 \big(\mb{H}^H\mb{H}+N_0\mb{I}_{K}\big)^{-1}\mb{H}^H\mb{y},
	 \end{eqnarray}
 which is then element-wise projected onto $\mc{S}$.  We note that the LMMSE filter in the presented form requires the inverse of a $K\times K$-dimensional matrix.
	
	\subsection{Level 3: Local Processing \& Large-Scale Fading Decoding}
	At this level, each AP pre-processes its received signal by computing a local estimate of $\mb{x}$ that are forwarded to the CPU for final decoding \cite{Emil2020Making}. Assuming full knowledge of channel matrix $\mb{H}_{\ell}$ at the $\ell$-th AP, the local LMMSE estimate $\check{\mb{x}}_{\ell} = [\check{x}_{i\ell},\ldots,\check{x}_{K\ell}]^T$ of $\mb{x}$ can be found as
	 \begin{eqnarray} 
	 	\check{\mb{x}}_{\ell} = \mb{H}_{\ell}^H \big(\mb{H}_{\ell}\mb{H}^H_{\ell}+N_0\mb{I}_{N}\big)^{-1}\mb{y}_{\ell}.
 	\end{eqnarray} 
	We note that the LMMSE filter in this presented form requires the inverse of a $N\times N$-dimensional matrix. The CPU then can linearly combine the local estimates $\{\check{x}_{i\ell}\,:\,\ell=1,\ldots,L\}$ to obtain the estimate
	 \begin{eqnarray} \label{linear-combination}
	 	\hat{x}_{i} = \sum_{\ell=1}^L a_{i\ell}\check{x}_{i\ell},
	 \end{eqnarray}
 	which is eventually used to decode $x_i$. Here, the weighting coefficient vector $\mb{a}_i = [a_{i1},\ldots,a_{iL}]^T$ relies only on channel statistics and can be optimized by the CPU. This combining method is also known as the large-scale fading decoding (LSFB) strategy in the context of cellular massive MIMO. We note that no instantaneous CSI of any channel is required at the CPU.
	 	
	\subsection{Level 2: Local Processing \& Simple Centralized Decoding}
	At this level, the CPU forms an estimate of $x_{i}$ by simply taking the average of the local estimates \cite{Emil2020Making}. This yield an estimate $\hat{x}_i$ as 
	 \begin{eqnarray} 
	 	\hat{x}_i = \frac{1}{L}\sum_{\ell=1}^{L}\check{x}_{i\ell}.
	 \end{eqnarray}
 	We note that no statistical parameters of CSI are needed at the CPU at this level of centralized signal processing.
	
	\subsection{Level 1: Small-Cell Network}
	At this level, each user signal is decoded by only one AP that gives the highest spectral efficiency to the user, i.e., the highest SINR \cite{Emil2020Making}. LMMSE filtering can be applied to obtain the local estimate of the user signal. Since only one estimate per user is forwarded to the CPU, no centralizing decoding is required.
	
\section{Variational Bayes for Cell-Free Detection}
 In this paper, we focus on developing VB-based methods for data detection in cell-free massive MIMO systems that require certain levels of centralized processing, i.e., Levels 4, 3, and 2. For Level 4 processing, we assume that the symbol vectors are estimated independently at each time slot. However, for Levels 3 and 2 processing,  we assume that the symbol vectors are first estimated locally over the whole fading block. As explained later in the section, this method of processing helps reduce the amount of signaling to the CPU, where the local estimates are aggregated to obtain the final estimate. 
	 
\subsection{Background on VB}
We first present the background on VB for approximate inference that will be exploited for solving the data detection in cell-free systems. VB inference is a powerful framework from machine learning that approximates intractable posterior distributions of latent variables with a known family of simpler distributions through optimization. The goal of VB inference is to find an approximation for a computationally intractable posterior distribution $p(\mb{x}|\mb{y})$ given a probabilistic model that specifies the joint distribution $p(\mb{x},\mb{y})$, where $\mb{y}$ represents the set of all observed variables and $\mb{x}$ is a set of $m$ latent variables and parameters. The VB inference method aims at finding a density function $q(\mb{x})$ with its own setting of variational parameters within a family $\mc{Q}$ of density functions that makes $q(\mb{x})$ close to the posterior distribution of interest $p(\mb{x}|\mb{y})$. VB inference amounts to solving the following optimization problem:
\begin{align}
	q(\mb{x}) &= \arg\min_{q(\mb{x}) \in \mc{Q}}\; \mr{KL}\big(q(\mb{x}) \|p(\mb{x}|\mb{y}) \big) \nonumber \\
	&= \arg\min_{q(\mb{x}) \in \mc{Q}}\;\mathbb{E}_{q(\mb{x})} \big[\ln q(\mb{x})\big] - \mathbb{E}_{q(\mb{x})}\big[\ln p(\mb{x}|\mb{y})\big] \; , 
\end{align}
where $\mr{KL}\big(q(\mb{x})\|p(\mb{x}|\mb{y})$ is the Kullback-Leibler (KL) divergence from $q(\mb{x})$ to $p(\mb{x}|\mb{y})$. Minimizing the KL divergence is equivalent to
maximizing the evidence lower bound ($\mr{ELBO}$)~\cite{bishop2006pattern}, which is defined as
\begin{align}
	\mr{ELBO}(q) = \mathbb{E}_{q(\mb{x})} \big[\ln p(\mb{x},\mb{y})\big] - \mathbb{E}_{q(\mb{x})} \big[\ln q(\mb{x}) \big] \; .
\end{align}
The maximum of $\mr{ELBO}(q)$ occurs when $q(\mb{x}) = p(\mb{x}|\mb{y})$. Since working with the true posterior distribution is often intractable, it is more convenient to consider a restricted family of distributions $q(\mb{x})$. Among VB inference methods, the \textit{mean-field approximation} enables efficient optimization of the variational distribution over a partition of the latent variables, while keeping the variational distributions over other partitions fixed~\cite{bishop2006pattern}. 
The mean-field variational family is constructed such that \begin{eqnarray} 
q(\mb{x}) = \prod_{i=1}^m q_i(x_i),
\end{eqnarray} 
where the latent variables are mutually independent and each is governed by a distinct factor in the variational density. Among all mean-field distributions $q(\mb{x})$, the general expression for the optimal solution of the variational density $q_i(x_i)$ that maximizes the ELBO can be obtained as~\cite{bishop2006pattern}
\begin{align}
	q_i(x_i) \propto \mr{exp}\left\{\big\langle{\ln p (\mb{y}|\mb{x}) + \ln p(\mb{x})\big\rangle}\right\} \; ,
\end{align}
where $\lr{\cdot}$ denotes the expectation with respect to all latent variables except $x_i$ using the currently fixed variational density $q_{-i}(\mb{x}_{-i}) = \prod_{j\neq i} q_{j}(x_{j})$. By iterating the update of $q_i(x_i)$ sequentially over all $j$, the $\mr{ELBO}(q)$ objective function can be monotonically improved. This is the basis behind the \textit{coordinate ascent variational inference} algorithm, which guarantees convergence to at least a local optimum of $\mr{ELBO}(q)$~\cite{bishop2006pattern,wainwright2008graphical}. To this send, we examine how the mean-field VB framework can be exploited for data detection at different levels of cooperation in a cell-free system.

\subsection{Level 4: Fully Centralized Processing}
At this level, the signals forwarded from the APs can be stacked into a single large-scale MIMO system as being shown in \eqref{eq_large_MIMO}. In a recent work \cite{Duy-MF-VB-2022}, we developed several VB-based methods for MIMO data detection. Among them, the \textbf{\textit{LMMSE-VB algorithm}} showed superior performance in MIMO systems with non-i.i.d. channels. Certainly, the algorithm can be adopted for data detection in cell-free systems with fully centralized processing. In the following, we present key operations in the algorithm. For details of the algorithm, we refer the readers to \cite{Duy-MF-VB-2022}.

The LMMSE-VB algorithm floats the background noise covariance matrix as an unknown random variable, instead of treating the noise's variance $N_0$ as known. The postulated noise covariance matrix $\mb{C}^{\mathrm{post}}$ is estimated by the algorithm itself. For ease of computation, we use $\mb{W} = (\mb{C}^{\mathrm{post}})^{-1}$ to denote the precision matrix and assume a conjugate prior complex Wishart distribution $\mc{CW}(\mb{W}_0,n)$ for $\mb{W}$, where $\mb{W}_0\succeq \mb{0}$ is the scale matrix and $n\geq NL$ indicates the degrees of freedom. The PDF of $\mb{W}\sim \mc{CW}(\mb{W}_0,n)$ satisfies 
\begin{eqnarray}
p(\mb{W}) \propto |\mb{W}|^{n-M}\mr{exp}\big(-\tr\{\mb{W}_0^{-1}\mb{W}\}\big).
\end{eqnarray} 
The joint distribution $p(\mb{y},\mb{x},\mb{W};\mb{H})$ can be factored as
\vspace{-0.1cm}
\begin{eqnarray}\label{factor-W}
p(\mb{y},\mb{x},\mb{W};\mb{H}) = p(\mb{y}|\mb{x},\mb{W};\mb{H})p(\mb{x})p(\mb{W}),
\end{eqnarray}
where $p(\mb{y}|\mb{x},\mb{W};\mb{H}) = \mc{CN}(\mb{y};\mb{Hx},\mb{W}^{-1})$. Given the observation $\mb{y}$, we aim at obtaining the mean-field variational distribution $q(\mb{x},\mb{W})$ such that
\begin{eqnarray}
p(\mb{x},\mb{W}|\mb{y};\mb{H}) \approx q(\mb{x},\mb{W}) = \prod_{i=1}^K q_i(x_i)q(\mb{W}).
\end{eqnarray}
The optimization of $q(\mb{x},\mb{W})$ is executed by iteratively updating $\{x_i\}$ and $\mb{W}$ as follows.

\textit{a) Updating $x_i$.} The variational distribution $q_i(x_i)$ is obtained by expanding the conditional in \eqref{factor-W} and taking the expectation with respect to all latent variables except $x_i$ using the variational distribution $\prod_{j\neq i}^K q_j(x_j)q(\mb{W})$:
\begin{eqnarray} \label{q-x-LMMSE-VB}
q_i(x_i)&\propto& p(x_i)\,\mc{CN}\big(z_i;x_i,{1}/{\big(\mb{h}_i^H\lr{\mb{W}}\mb{h}_i\big)} \big),
\end{eqnarray}
where $z_i$ is a linear estimate of $x_i$ that is defined as
\begin{eqnarray} \label{z-i-LMMSE-VB}
z_i	&=& \lr{x_i} + \frac{\mb{h}^H_i\lr{\mb{W}}}{\mb{h}_i^H\lr{\mb{W}}\mb{h}_i}\big(\mb{y} - \mb{H}\lr{\mb{x}}\big).
\end{eqnarray}

It is observed in \eqref{q-x-LMMSE-VB} that $\mc{CN}\big(z_i;x_i,\hat{\sigma}_i^2\big)$ with $\hat{\sigma}_i^2 =1/\big(\mb{h}_i^H\lr{\mb{W}}\mb{h}_i\big)$ can be interpreted as the likelihood function $p\big(z_i|x_i;\hat{\sigma}_i^2\big)$. In other words, the mean-field VB approximation decouples the linear MIMO system into $K$ parallel AWGN channels $z_i = x_i + \mc{CN}\big(0,\hat{\sigma}_i^2\big)$.
	
The variational distribution $q_i(x_i)$ is realized by normalizing $p(x_i)\,\mc{CN}\big(z_i;x_i,\hat{\sigma}_i^2\big)$. The variational mean $\lr{x_i} = \mathbb{E}[x_i|z_i]$ and variance $\sigma_{x_i}^2$ are then computed  accordingly. 


\textit{b) Updating $\mb{W}$.} The variational distribution $q(\mb{W})$ is obtained by taking the expectation of the conditional in \eqref{factor-W} with respect to $q(\mb{x})$:
\begin{eqnarray} \label{q-W}
q(\mb{W}) &\propto& \mr{exp}\big\{\big\langle\ln p(\mb{y}|\mb{x},\mb{W};\mb{H}) + \ln p(\mb{W}) \big\rangle\big\}.
\end{eqnarray}
The variational distribution $q(\mb{W})$ is also complex Wishart with $n+1$ degrees of freedom \cite{Duy-MF-VB-2022}. The variational mean $\lr{\mb{W}}$ can be computed accordingly. In \cite{Duy-MF-VB-2022}, we also proposed to use the estimator 
\begin{eqnarray}
\lr{\mb{W}} = \bigg(\frac{\|\mb{y}-\mb{Hx}\|^2}{NL}\mb{I}_{NL} + \mb{H}\bs{\Sigma}_{\mb{x}}\mb{H}\bigg)^{-1},
\end{eqnarray}
where $\bs{\Sigma}_{\mb{x}} = \mr{diag}(\sigma_{x_1}^2,\ldots,\sigma_{x_K}^2)$. 

By iteratively optimizing $\big\{q_i(x_i)\big\}$ and $q(\mb{W})$ via the updates of $\{\lr{x_i}\}$ and $\lr{\mb{W}}$, we obtain the CAVI algorithm for estimating $\mb{x}$ and the precision matrix $\mb{W}$. We refer to this scheme as the LMMSE-VB algorithm since $z_i$ resembles an LMMSE estimate of $x_i$ due to the cancellation of the inter-user interference and the whitening with the postulated noise covariance matrix $\mb{C}^{\mathrm{post}}$.

\subsection{Level 3: Local Processing \& Nonlinear Decoding}
At this level, our proposed VB-based method involves two operations: 1) Executing the LMMSE-VB algorithm independently at each AP to compute local estimates of $\mb{x}_t$ and 2) Aggregating the local estimates at the CPU for joint nonlinear decoding of $\mb{x}_t$. However, we make a minor modification to the LMMSE-VB algorithm which allow it to operate over the whole block of $T$ time slots.

\subsubsection{AP Processing}
The signal processing at an AP, say the $\ell$-th AP, is to generate a coarse estimate $\hat{\mb{x}}_t$ of $\mb{x}_t$, from the observation $\mb{y}_t$. We treat the background noise covariance matrix at the $\ell$-th AP as an unknown random variable. The postulated noise matrix $\mb{C}_{\ell}^{\mr{post}}$ has to be estimated as well. We denote the precision matrix $\mb{W}_{\ell}=(\mb{C}_{\ell}^{\mr{post}})^{-1}$,  $\mb{Y}_{\ell}=[\mb{y}_{\ell,1},\ldots,\mb{y}_{\ell,T}]$, and $\mb{X} = [\mb{x}_{1},\ldots,\mb{x}_{T}]$. The joint distribution $p(\mb{Y}_{\ell}, \mb{X}, \mb{W}_{\ell}; \mb{H}_{\ell})$ can be factorized as
\begin{equation}\label{factor-W-l}
p(\mb{Y}_{\ell}, \mb{X}, \mb{W}_{\ell}; \mb{H}_{\ell}) = p(\mb{Y}_{\ell} |\mb{X}, \mb{W}_{\ell}; \mb{H}_{\ell})p(\mb{X})p(\mb{W}_{\ell}),
\end{equation}
where $p(\mb{Y}_{\ell} |\mb{X}, \mb{W}_{\ell}; \mb{H}_{\ell}) = \prod_{t=1}^Tp(\mb{y}_{\ell,t}|\mb{x}_t,\mb{W}_{\ell};\mb{H}_{\ell})$ with  $p(\mb{y}_{\ell,t}|\mb{x}_t,\mb{W}_{\ell};\mb{H}_{\ell})= \mc{CN}\big(\mb{y}_{\ell,t};\mb{H}_{\ell}\mb{x}_t,\mb{W}_{\ell}^{-1}\big)$. 
Given the observation $\mb{Y}_\ell$, we aim at obtaining the mean-field variational distribution $q_\ell(\mb{X},\mb{W}_\ell)$ such that
\begin{align}
p(\mb{X},\mb{W}_{\ell}|\mb{Y}_\ell;\mb{H}_\ell) &\approx q_\ell(\mb{X},\mb{W}_\ell) \nonumber\\
&= \prod_{i=1}^K\prod_{t=1}^T q_{i\ell,t}(x_{i,t})q(\mb{W}_\ell).
\end{align}
The optimization of $q_\ell(\mb{X},\mb{W}_\ell)$ is executed by iteratively updating $\{x_{i,t}\}$ and $\mb{W}_\ell$ as follows.

\textit{a) Update $x_{i,t}$:} The variational distribution $q_{i\ell,t}(x_{i,t})$ is obtained by expanding the conditional in \eqref{factor-W-l} and taking the expectation with respect to all latent variables except $x_{i,t}$ using the variational distribution $\prod_{(j,r)\neq(i,t)} q_{j\ell,r}(x_{j,r})q(\mb{W}_\ell)$:
\begin{align}\label{q-x-local}
&q_{i\ell,t}(x_{i,t}) \nonumber \\
&\propto \exp \left \{\langle \ln p(\mb{y}_{\ell,t}|\mb{x}_t,\mb{W}_{\ell};\mb{H}_{\ell}) + \ln p(\mb{x}_t) \rangle \right \} \notag \\ 
&\propto p(x_{i,t})\exp \left \{\left \langle -(\mb{y}_{\ell,t}-\mb{H}_{\ell}\mb{x}_t)^H\mb{W}_{\ell}(\mb{y}_{\ell,t}-\mb{H}_{\ell}\mb{x}_t) \right \rangle \right\} \notag\\
&\propto p(x_{i,t})\exp \left \{ -\mb{h}_{i\ell}^H\langle \mb{W}_{\ell} \rangle\mb{h}_{i\ell} |x_{i,t} - z_{i\ell,t}|^2\right \} \notag\\
&\propto p(x_{i,t})\,\mc{CN}\big(z_{i\ell,t};x_{i,t},1/(\mb{h}_{i\ell}^H\langle \mb{W}_{\ell} \rangle\mb{h}_{i\ell})\big),
\end{align}
where 
\begin{eqnarray}
z_{i\ell,t} &=& \frac{\mb{h}_{i\ell}^H\langle \mb{W}_{\ell} \rangle}{\mb{h}_{i\ell}^H\langle \mb{W}_{\ell} \rangle\mb{h}_{i\ell}} \big(\mb{y}_{\ell,t} - \sum_{j\neq i}^K \mb{h}_{j\ell}\langle x_{j\ell,t}\rangle \big) \nonumber \\
&=& \langle x_{i,t}\rangle + \frac{\mb{h}_{i\ell}^H\langle \mb{W}_{\ell} \rangle(\mb{y}_{\ell,t} - \mb{H}_{\ell}\langle\mb{x}_{t}\rangle)}{\mb{h}_{i\ell}^H\langle \mb{W}_{\ell} \rangle\mb{h}_{i\ell}}.
\end{eqnarray}

It is observed in \eqref{q-x-local} that $\mc{CN}\big(z_{i\ell,t};x_{i,t},\check{\sigma}_{i\ell}^2\big)$ with $\check{\sigma}_{i\ell}^2={1}/{\big(\mb{h}_{i\ell}^H\lr{\mb{W}_\ell}\mb{h}_{i\ell}\big)}$  can be interpreted as the likelihood function $p\big(z_{i\ell,t}|x_{i,t};\check{\sigma}_{i\ell}^2\big)$. In this case, the mean-field VB approximation decouples the uplink MIMO channel to the $\ell$-th AP into $K$ parallel AWGN channels $z_{i\ell,t} = x_{i,t} + \mc{CN}\big(0,\check{\sigma}_{i\ell}^2\big)$. It is also observed that $z_{i\ell,t}$ is the local LMMSE estimate of $x_{i,t}$, while the variance $\check{\sigma}_{i\ell}^2$ indicates the reliability of this estimate. 

The variational distribution $q_{i\ell,t}(x_{i,t})$ is realized by normalizing $p(x_{i,t})\mc{CN}\big(z_{i\ell,t};x_{i,t},\check{\sigma}_{i\ell}^2\big)$. The variational mean $\lr{x_{i,t}} = \mathbb{E}[x_{i,t}|z_{i\ell,t}]$ and variance $\sigma_{x_{i,t}}^2$ can be computed accordingly. Hereafter, we use $\check{x}_{i\ell,t}$ instead of $\lr{x_{i,t}}$ or $\mathbb{E}[x_{i,t}|z_{i\ell,t}]$ to indicate the nonlinear MMSE estimate of $x_{i,t}$ at the $\ell$-th AP.

\textit{b) Update $\mb{W}_{\ell}$:} The variational distribution $q(\mb{W}_\ell)$ is obtained by taking the expectation of the conditional in \eqref{factor-W-l} with respect to $\prod_{i=1}^K\prod_{t=1}^T q_{i\ell,t}(x_{i,t})$:
\begin{align} \label{q-W-2}
q(\mb{W}_\ell) &\propto& \mr{exp}\big\{\big\langle\ln p(\mb{Y}_{\ell}|\mb{X},\mb{W}_\ell;\mb{H}_\ell) + \ln p(\mb{W}_\ell) \big\rangle\big\}.
\end{align}
\begin{figure*}
\begin{eqnarray}\label{W-l}
\lr{\mb{W}_{\ell}} = (n+T) \Bigg(\mb{W}_0 + (\mb{Y}_{\ell}-\mb{H}_\ell\mb{X})(\mb{Y}_{\ell}-\mb{H}_\ell\mb{X})^H + \sum_{t=1}^T\mb{H}_\ell\bs{\Sigma}_{\mb{x},t}\mb{H}_\ell\Bigg)^{-1}.
\end{eqnarray}
\hrulefill
\end{figure*}
Assuming a conjugate prior complex Wishart distributed $\mc{CW}(\mb{W}_{0,\ell},n)$ for $\mb{W}_\ell$, the variational distribution $q(\mb{W})$ is also complex Wishart with $n+T$ degrees of freedom. The variational mean $\lr{\mb{W}_\ell}$ is given in \eqref{W-l}, where $\bs{\Sigma}_{\mb{x},t} = \diag(\sigma_{x_{1,t}}^2, \ldots,\sigma_{x_{K,t}}^2)$.

The LMMSE-VB algorithm is executed at the $\ell$-th AP by iteratively optimizing $\{q_{i\ell,t}(x_{i,t})\}$ and $q(\mb{W})$ via the updates of $\{\lr{x_{i,t}}\}$ and $\lr{\mb{W}_\ell}$. The $\ell$-th AP then sends the LMMSE estimate $z_{i\ell,t}$ and the variance $\check{\sigma}_{i\ell}^2$ to the CPU for centralized decoding. By pre-processing the whole block of $T$ time slots, $\check{\sigma}_{i\ell}^2$ is sent only once for each channel realization. In contrast, if the LMMSE-VB algorithm is executed on a per time slot basis, the variance of the LMMSE estimate $z_{i\ell,t}$ has to be computed and sent for each time slot.

\subsubsection{CPU Processing}
After collecting the local estimates $z_{i\ell,t}$ and the variance $\check{\sigma}_{i\ell}^2$ from the $L$ APs, the CPU can proceed to decode each of the $K$ symbols independently. Since $z_{i\ell,t} = x_{i,t} + \mc{CN}\big(0,\check{\sigma}_{i\ell}^2\big)$, an approximate posterior distribution $p(x_{i,t}|\{z_{i\ell,t}\};\{\check{\sigma}_{i\ell}^2\})$ can be easily derived. The MAP estimate $\hat{x}_{i,t}$ of $x_{i,t}$ is obtained as
\begin{equation}\label{nonlinear-combination}
\hat{x}_{i,t} = \arg\max_{x_{i,t} \in \mc{S}} \left(\ln p(x_i)-\sum_{\ell=1}^L \frac{|z_{i\ell,t} - x_{i,t}|^2}{\check{\sigma}_{i\ell}^2}\right).
\end{equation}

We note that the above nonlinear combination of local estimates and reliability information is significantly different from the linear combination of local estimates in \eqref{linear-combination}. 


\subsection{Level 2: Local Processing \& Simple Linear Combining}
At this level, only local estimates are fed back to the CPU. The LMMSE-VB mentioned in Level 3 signal processing can be used to generate the coarse local estimates. However, the local nonlinear MMSE estimates $\check{x}_{i\ell,t}$ is sent, instead of the LMMSE estimate $z_{i\ell,t}$ and the variance $\check{\sigma}_{i\ell}^2$. We note that $\check{x}_{i\ell,t}$ can be computed using $z_{i\ell,t}$ and $\check{\sigma}_{i\ell}^2$, but not the reverse. 

A simple estimate of $x_{i,t}$ can be obtained by simply taking the average of all the estimates $\check{x}_{i\ell,t}$ as
\begin{equation}
\hat{x}_{i,t} = \frac{1}{L}\sum_{\ell=1}^L \check{x}_{i\ell,t}.
\end{equation}
The final detected symbol of $x_{i,t}$ is the constellation point that is closest to $\hat{x}_{i,t}$.

\section{Numerical Results}
\label{sec_numerical_results}
This section presents the numerical results comparing the developed VB-based methods for data detection in cell-free systems with the LMMSE filtering methods in \cite{Emil2020Making}. We use a simulation setting and a channel model in urban environments similar to the work in~\cite{Emil2020Making}. In particular, a network area of $1 \times 1$ km is considered where the APs are deployed on a square grid and users are randomly distributed. The large-scale fading coefficient of the channel between user-$i$ and AP-$\ell$ (in dB) is given as
\begin{equation}
    \beta_{i\ell} = -30.5 - 36.7\log_{10}(d_{i\ell}) + F_{i\ell},
\end{equation}
where $d_{i\ell}$ (in m) is the distance between user-$i$ and AP-$\ell$ and $F_{i\ell}\sim\mathcal{N}(0,16)$ is the shadow fading. The correlation between the shadowing terms from an AP to different users is modeled as
\begin{equation}
    \mbb{E}[F_{i\ell}F_{i'\ell'}] = \begin{cases}
    16\times2^{-\delta_{ii'}/9}, & \ell = \ell' \\
    0, & \ell \neq \ell'
    \end{cases}
\end{equation}
where $\delta_{ii'}$ (in m) is the distance between user-$i$ and user-$i'$. Receive antennas at each AP are arranged in a uniform linear array with half-wavelength spacing. For spatial correlation, we use the Gaussian local scattering model with a $15^{\circ}$ angular standard deviation~\cite{bjornson2017massive}. We set the noise as $\mathcal{CN}(0,1)$ and vary the transmit power of users. 

\begin{figure}
	\centering
	\includegraphics[width=1\linewidth]{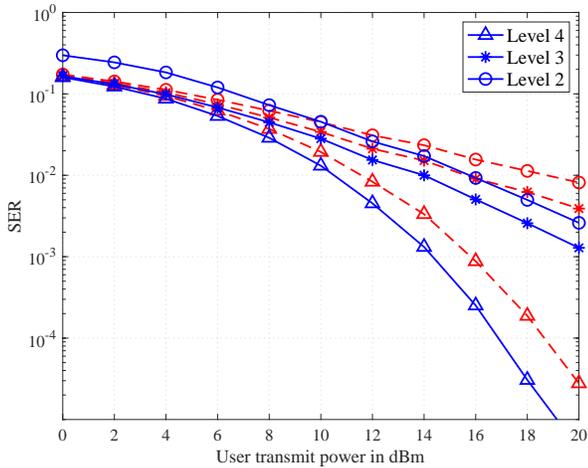}
	\caption{SER performance of the VB-based methods (in \emph{solid} lines) and LMMSE methods (in \emph{dashed} lines) \emph{versus} the user transmit power, with $K=16$, $L=16$, and $N=4$.}
	\label{fig_1}
\end{figure}

In this work, we compare different data detection methods assuming perfect CSI and QPSK signalling. We assume that each AP is equipped with $4$ antennas, i.e., $N=4$.
Fig.~\ref{fig_1} presents the symbol error rate (SER) performance of the two types of methods in a relatively small setting of cell-free systems with $K=16$ and $L=16$. As the user transmit power is increased, the VB-based methods attain much lower SER than the MMSE filtering methods. Up to $2$-dB gain is observed at Level 4 and $4$-dB gain is observed at Level~3 and~2. 

\begin{figure}
	\centering
	\includegraphics[width=1\linewidth]{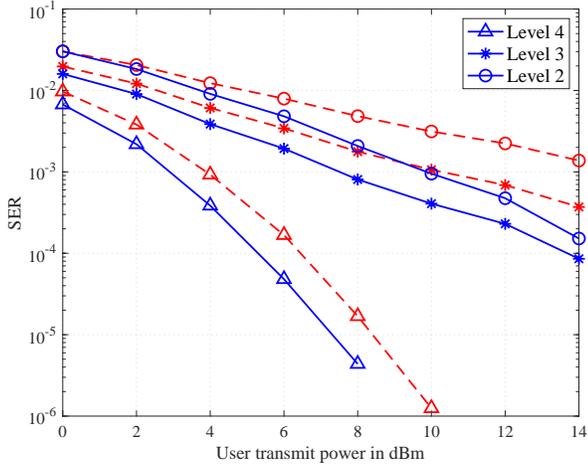}
	\caption{SER performance of the VB-based methods (in \emph{solid} lines) and LMMSE methods (in \emph{dashed} lines) \emph{versus} the user transmit power, with $K=40$, $L=64$, and $N=4$.}
	\label{fig_2}
\end{figure}

Fig. \ref{fig_2} presents the SER performance a cell-free system with $K=40$ and $L=64$. The figure clearly indicates the superior performance of the proposed VB-based methods over  the MMSE filtering methods. It is also observed from both figures that the more centralized signal processing is carried at the CPU, the better SER performance can be achieved, especially in systems with a large number of users, e.g., $K=40$. 

\section{Conclusion}
In this paper, we have proposed the VB-based methods for data detection in cell-free systems at three different levels of AP cooperation. The proposed methods can achieve much lower SER than the linear MMSE signal processing methods. We note that the presented study only considers the case of perfect CSI available at the CPU (for Level 4) and at the APs (for Levels 3 and 2). As an extension of this paper, we are developing novel VB-based methods for data detection with imperfect CSI and joint channel estimation and data detection in cell-free systems. 

\label{sec_conclusion}

\ifCLASSOPTIONcaptionsoff
\newpage
\fi

\section*{Acknowledgment}
This work was supported by the U.S. National Science Foundation under Grants ECCS-2146436 and CCF-2225576.

\bibliographystyle{IEEEtran}
\bibliography{ref}

	
	%
	
	
	
	
	
	
	
	

\end{document}